\documentclass[10pt,twocolumn]{IEEEtran}

\makeatletter
\def\ifundefined{\@ifundefined}
\makeatother

\usepackage{txfonts}
\usepackage{graphicx}
\def\U#1{{%
\def\O{\mbox{O}}
\def\u{\mbox{u}}
\mathcode`\u=\mu
\mathcode`\O=\Omega
\mathrm{#1}}}
\def\ii{{\mathrm{i}}}
\def\ee{{\mathrm{e}}}
\def\dd{{\mathrm{d}}}
\def\sub#1{_{\mbox{{\scriptsize#1}}}}
\def\gtrsim%
     {\mathrel{\raisebox{0.4ex}{\hbox{$<$}}\kern-0.75em
    \raisebox{-0.5ex}{\hbox{$\sim$}}}}
\def\lesssim%
     {\mathrel{\raisebox{0.4ex}{\hbox{$>$}}\kern-0.75em
    \raisebox{-0.5ex}{\hbox{$\sim$}}}}
\begin{document}
\title{Negative Group Delay and Superluminal Propagation:
An Electronic Circuit Approach}
\author{M. Kitano, T. Nakanishi, and K. Sugiyama
\thanks{
This work was supported in part by the Ministry of
Education, Culture, Sports, Science and Technology
in Japan under a Grant-in-Aid for Scientific Research
No.~11216203
and by the National Science Foundation.}
\thanks{
M. Kitano, T. Nakanishi, and
K. Sugiyama are with
the Department of Electronic Science and Engineering,
Kyoto University, Kyoto 606-8501 JAPAN
(e-mail: kitano@kuee.kyoto-u.ac.jp;
t-naka@giga.kuee.kyoto-u.ac.jp;
sugiyama@kuee.kyoto-u.ac.jp).
}
\thanks{
This paper will appear in
the January/February 2003 issue of
the IEEE {\sl Journal of Selected Topics in Quantum Electronics}.
}
}

%
%

\maketitle

\begin{abstract}
We present a simple electronic circuit which provides
negative group delays for band-limited, base-band pulses.
It is shown that large time advancement comparable to the pulse width
can be achieved with appropriate cascading of negative-delay circuits but
eventually the out-of-band gain limits the number of cascading.
The relations to superluminality and causality are also discussed.
\end{abstract}


\section{Introduction}
Brillouin and Sommerfeld showed that
in the region of anomalous dispersion,
which is inside of the absorption band,
the group velocity can exceed $c$,
the light speed in a vacuum, or even
be negative \cite{Brillouin,Chu}.
Recently, it was shown that for a gain medium,
superluminal propagation is possible
at the outside of the gain resonance.
Superluminal effects are also predicted
in terms of quantum tunneling or evanescent waves
\cite{Chiao,Enders,Steinberg}.
Superluminal group velocities have been confirmed experimentally
in various systems, 
and most controversies over
this counterintuitive phenomenon have settled down.
However, there seem a several questions remain open;
for example, 
``How far we can speed up the wave packets,''
``Is it really nothing to do with information transmission,''
``What kind of applications are possible,'' and so on.
In this paper we will try to solve some of these problems
by utilizing a simple circuit model for negative group delays.

Negative delay in lumped systems such as electronic circuits
is very helpful to understand various aspects of
superluminal group velocity.
Mitchell and Chiao \cite{Mitchell,MitchellPL} constructed a
bandpass amplifier with an $LC$ resonator and an operational
amplifier.
An arbitrary waveform generator is used to
generate a gaussian pulse by which a carrier
is modulated.
The circuit basically emulates an optical gain medium which
shows anomalous dispersion in off-resonant region.
Wang {\it et al.} \cite {Cao} extended this circuit 
by using two $LC$ resonators which correspond to 
the two Raman gain lines \cite{Wang,Dogariu}.
At the middle of two gain peaks the frequency dependence
of amplitude response is compensated and the pulse distortion
can be minimized.

The present authors \cite{Nakanishi}
used an operational amplifier with 
an $RC$ feedback circuit.
It provides negative delays for baseband pulses.
In previous experiments, optical or electronic, a carrier 
frequency ($\omega_0$) is modulated by a pulse which varies
slowly compared with the carrier oscillation
and the displacement of the envelopes is measured.
Without carriers ($\omega_0=0$), the system becomes much
more simple.
The amplitude response symmetric with respect to zero frequency
is helpful to reduce the distortion.
The baseband pulse is simply derived 
from a rectangular pulse generator
and a series of lowpass filters.

The time constants can easily be set at the order of seconds and we
can actually observe that the output LED (light-emitting diode) 
is lit earlier than the input LED.
In addition to the usefulness as a demonstration tool,
this circuit turned out to be very convenient to
look into the essentials of negative group delays and
superluminal propagation because of its simplicity.

In this paper we exploit the circuit model
in order to investigate some of the fundamental problems.
First we discuss the relation between negative group delay and
superluminarity, and then the
approximate realization of (positive and negative) delays
by lumped systems.
Then we consider the spectral condition imposed on input pulses
and describe the design of lowpass filters for pulse preparation.
Next in order to increase the advancement, a number of
negative delay circuits are cascaded.
We find that an advancement as large as the pulse width is possible
but the slow increase of the advancement and the exponential increase of
out-of-band gain almost prohibit the achievement of further advancements.
Finally by regarding our system as a communication channel,
we discuss the causality in lumped systems.

\section{Negative delay and superluminal propagation}

The group velocity $v\sub{g}$ in a dispersive medium is defined as
\begin{equation}
v\sub{g}^{-1}=\left.\frac{\dd k}{\dd \omega}\right|_{\omega_0}
,
\end{equation}
where the wavenumber $k(\omega)$ is a function of frequency $\omega$.
It corresponds to the propagation speed of an envelope of
signal whose spectrum is limited within a short interval containing
$\omega_0$.

Similarly the group delay is defined as
\begin{equation}
t\sub{d}=-\left.\frac{\dd\phi}{\dd\omega}\right|_{\omega_0}
,
\end{equation}
where $\phi(\omega)$ represents the frequency-dependent phase
shift.
It corresponds to the temporal shift of the envelope of the
band-limited signal passing through a system.
For a medium with length $L$,
the phase shift is given by
$
\phi(\omega)=-k(\omega)L
$
and we have
\begin{equation}
t\sub{d}=\frac{\dd (kL)}{\dd \omega}=\frac{\dd k}{\dd\omega}L 
= v\sub{g}^{-1}L.
\end{equation}

These two quantities seem almost identical,
but the group delay is a more general concept because 
it can be defined even for a lumped system.
The lumped system is a system whose size $L$ is
much smaller than the wavelength $2\pi/k$ of interest.
Neglecting the propagation effects,
the behavior of the system can be described by a set of ordinary
differential equations with respect to time.
For distributed systems, on the other hand,
the spatio-temporal 
partial differential equation must be used.

\begin{figure}
\begin{center}
\includegraphics[scale=0.8]{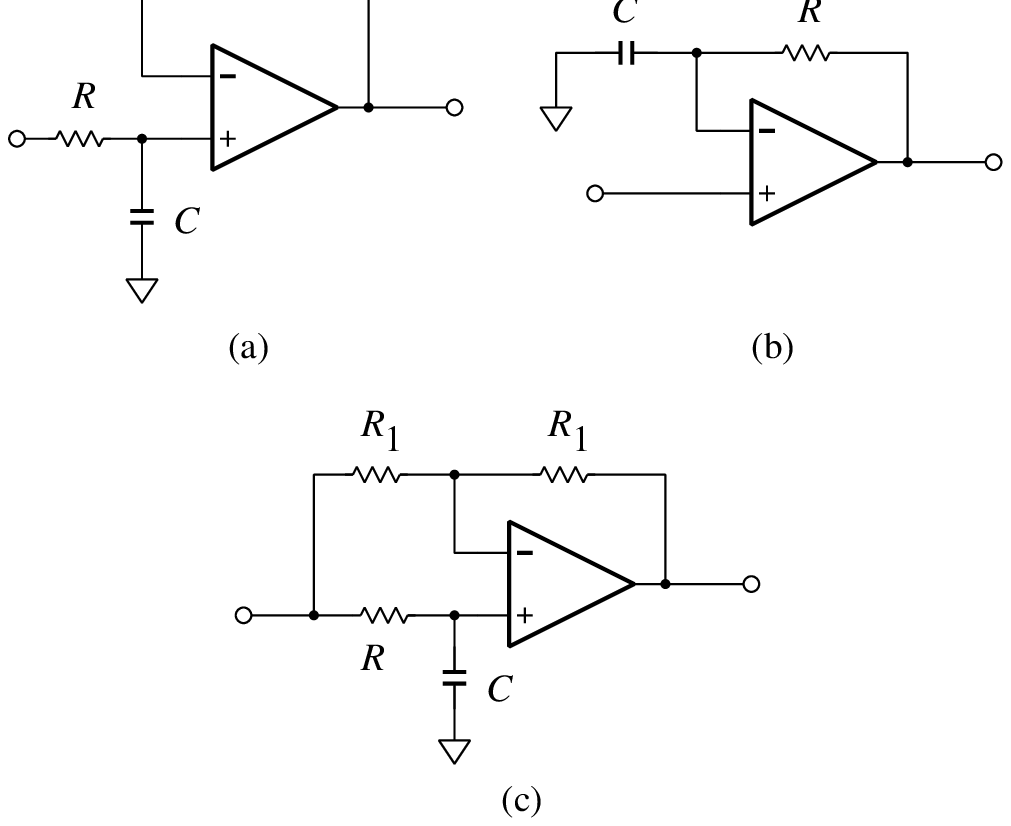}
\end{center}
\caption{
Elementary circuits used in 
negative group delay experiments.
(a) A first order low-pass filter. 
A series of low-pass filters are used for pulse preparation.
(b) Negative group delay circuit.
This simple circuit provides time-advancement for baseband pulses.
(c) All-pass filter.
}
\label{fig:elementary_cir}
\end{figure}

The relation between negative group delays and superluminality
can easily be understood when we consider a system
consisting with a vacuum path (length $L$) and a lumped
system (delay $t\sub{d}$) which is located at the end of
the path.
The total time $t\sub{total}$ required for a pulse to
pass through the system is
\begin{equation}
t\sub{total} = \frac{L}{c}+t\sub{d}
.
\label{eq:ttotal}
\end{equation}
The corresponding velocity  $v\sub{g}=L/t\sub{total}$ satisfies the relation:
\begin{equation}
 \frac{1}{v\sub{g}}=\frac{1}{c}+\frac{t\sub{d}}{L}
.
\label{eq:vgdecom}
\end{equation}
For $t\sub{d}>0$ (positive delay), $v\sub{g}$ is smaller than $c$.
For $t\sub{d}<0$ (negative delay), there are two cases.
In the case $(-t\sub{d})< L/c$, $v\sub{g}$ is larger than $c$
(superluminal in a narrow sense),
while in the case $(-t\sub{d})> L/c$, $v\sub{g}$ becomes negative
(negative group velocity).
In the latter case the contribution of the lumped part dominates
that of the free propagation path.

Normally the superluminality has been considered as a propagation
effect.
But in many cases, it seems more appropriate to discuss
in term of the negative group delay for a lumped system.
Let us take experimental parameters from \cite{Wang},
in which the negative delay of $t\sub{d}=-60\,\U{ns}$ was observed.
We note that the pulse length
$cT\sub{w}=c\times 2\,\U{us}=600\,\U{m}$
is much longer than the cell length $L=6\,\U{cm}$.
Therefore we can safely use the lumped approximation.
Since we can eliminate the carrier frequency by the slowly-varying
envelope approximation, the wavelength of light ($\sim \U{um}$)
will not come into play any more.

It should be stressed that the cases where
the second term in Eqs. (\ref{eq:ttotal}) or (\ref{eq:vgdecom}), 
which is positive
or negative, dominates the first term are very likely.
For typical atomic experiments,
the bandwidth $\Delta\omega$, which is of the order of
MHz to GHz, roughly determines advancement as
$|t\sub{d}|\sim \Delta nk_0 L/\Delta\omega\sim 1/\Delta\omega$,
$\U{us}$ to $\U{ns}$, while the passage time $L/c$ is
less than the order of$\U{ns}$.
Forcible assignment of a velocity to such cases,
$v\sub{g}=-(1/300)c$ in the above example, would have caused 
some confusions.

\section{Group delays -- ideal and approximate}
\subsection{Mathematical representation}

A mathematical representation for ideal delays
can be written as
\begin{equation}
v\sub{out}(t) = (h * v\sub{in}) (t)
= v\sub{in}(t-t\sub{d})
,
\end{equation}
where
$h(t)=\delta(t-t\sub{d})$
is the impulse response of the system and
$t\sub{d}$ is the delay time.
Its Fourier transform is given by
\begin{equation}
\tilde{V}\sub{out}(\omega) 
= H\sub{D}(\omega)\tilde{V}\sub{in}(\omega)
,
\end{equation}
with
\begin{equation}
 H\sub{D}(\omega) \equiv  
\int_{-\infty}^{\infty} \dd t\, h(t)\ee^{-\ii\omega t}=
\exp (-\ii\omega t\sub{d})
.
\end{equation}

For $t\sub{d}>0$ (positive delay),
the impulse response $h(t)$ is causal, i.e.,
it is zero for $t<0$.
The positive delay can be realized easily,
if you have an appropriate space ($L=ct\sub{d}$).
But
there is no way to make ideal, unconditional negative delays,
because $h(t)$ is non-causal in this case.

\subsection{Building blocks}

It should be noted that
no lumped systems ($L=0$) can produce ideal positive or negative delays.
From now on let us consider how to 
make approximate delays with lumped systems. 
The amplitude and the phase of the ideal response function $H\sub{D}(\omega)$
are
\begin{eqnarray}
A\sub{D}(\omega) &\equiv& |H\sub{D}(\omega)| = 1,\\
\phi\sub{D}(\omega) &\equiv& \mbox{arg}\,H\sub{D}(\omega) = -t\sub{d}\omega
.
\end{eqnarray}
In \cite{Nakanishi}, 
we used a circuit having a transfer function
\begin{equation}
H(\omega)=1+\ii\omega T,
\end{equation}
by which negative delay is provided for baseband pulses ($\omega_0=0$).
Here we will examine several transfer functions which can be
realized with an operational amplifier and a few passive components.

First, we consider a function with a single pole:
\begin{eqnarray}
H\sub{L}(\omega)&=&\frac{1}{1+\ii\omega T}
,\\
A\sub{L}(\omega)&=&1/\sqrt{1+(\omega T)^2} \sim
 1 - \frac{(\omega T)^2}{2}
,\\
\phi\sub{L}(\omega)&=& -\tan^{-1}\omega T \sim - T \omega 
.
\label{eq:lowpass}
\end{eqnarray}
The stability condition
that all the poles reside in the upper half plane
requires $T>0$,
therefore, only positive delays $t\sub{d}=T>0$ can be achieved
with this type of transfer function.
An example of circuit is shown in 
Fig.~\ref{fig:elementary_cir}(a).
Only in the
region $|\omega| < 1/T$, 
the amplitude response is flat and 
the phase response is linear. 
The circuit works only for band-limited signals.

Secondly, we will check a function with a single zero:
\begin{eqnarray}
H(\omega)&=&1+\ii\omega T
,\\
A(\omega)&=&\sqrt{1+(\omega T)^2} \sim 1 + \frac{(\omega T)^2}{2}
\label{eq:hamp},\\
\phi(\omega)&=&\tan^{-1}\omega T \sim T\omega
\label{eq:hphase}.
\label{eq:highpass}
\end{eqnarray}
In this case, no sign restrictions are imposed on $T$,
therefore, both positive and negative delays can be realized;
$t\sub{d}= -T$.
A circuit for $H(\omega)$ ($T>0$) is shown in
Fig.~\ref{fig:elementary_cir}(b).
Perhaps this is the most simple circuit which provides negative delay.
Again it works only in the region $|\omega| < 1/T$. 
Even worse is the rising of gain $A(\omega)$ at the outside of
the band.
We can also construct a positive delay circuit utilizing the
relation $2-H(\omega)=1+\ii\omega(-T)$.

By observing the sign restrictions for 
$H\sub{L}(\omega)$ and $H(\omega)$,
we notice that
an asymmetry between the positive and negative delays exists even
in lumped systems.

Another interesting transfer function is
\begin{eqnarray}
H\sub{A}(\omega)&=&\frac{1-\ii\omega T}{1+\ii\omega T}
,\\
A\sub{A}(\omega)&=&1
,\\
 \phi\sub{A}(\omega)&=&-2\tan^{-1}\omega T \sim -2T\omega
,
\label{eq:allpass}
\end{eqnarray}
which can be realized by the circuit shown in
Fig.~\ref{fig:elementary_cir}(c).
This circuit is called the all-pass filter.
The phase function is the same as the above cases
aside from the factor 2, but
the amplitude response in independent of the frequency
as in the case of ideal delay.
The stability condition implies $T>0$, therefore,
only positive delays are possible.

\section{Bandwidth and distortion}
\subsection{Bandwidth of negative delay circuit}

It turned out that lumped circuits can provide a 
delay, positive or negative, only for
a band-limited signal.
From the approximate transfer function,
$H(\omega) = 1+\ii \omega T$, for negative delays,
we have the imperfect amplitude and phase functions:
\begin{figure}
\begin{center}
\includegraphics[scale=0.8]{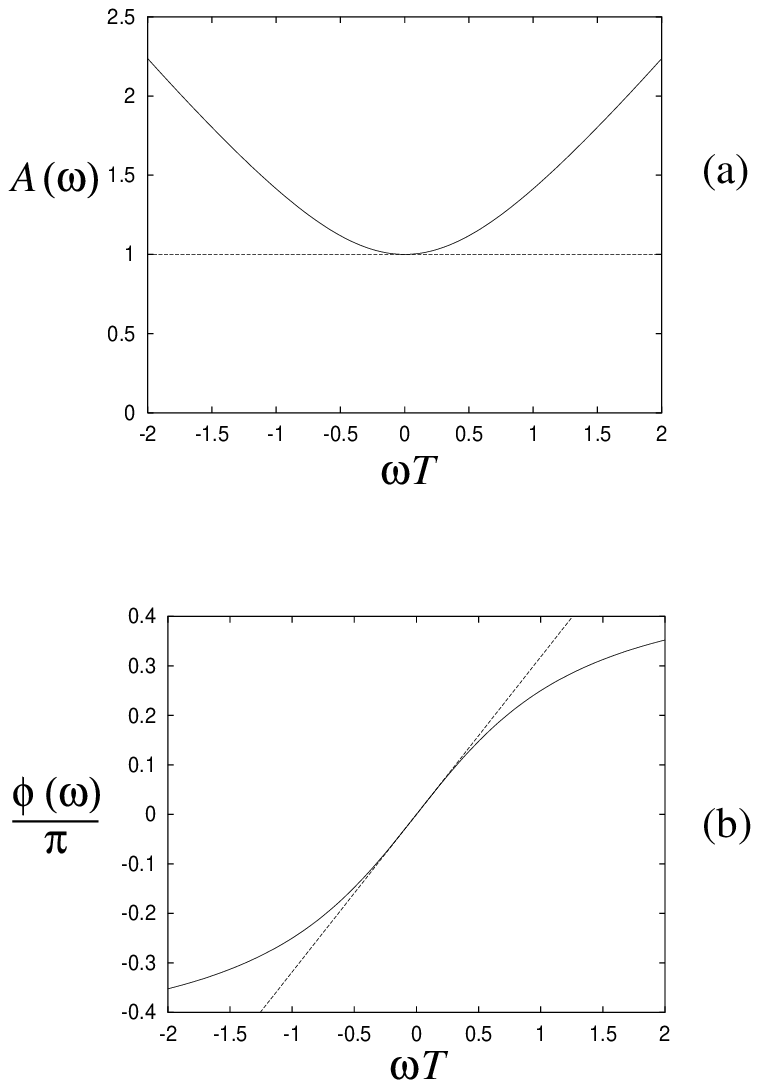}
\end{center}
\caption{Frequency response of approximate negative
delay; (a) amplitude, (b) phase responses.
The response for ideal negative delay are plotted with
dashed lines.
The approximation is valid only for $|\omega|<1/T$.
}
\label{fig:freq_response}
\end{figure}
\begin{eqnarray}
A(\omega)&=&1+O(\omega^2 T^2)
\label{eq:amp_imp}
,\\
\phi(\omega)&=&\omega T + O(\omega^3 T^3)
\label{eq:ph_imp}
,
\end{eqnarray}
which are plotted in
Fig.~\ref{fig:freq_response}.
We see that the inputs must satisfy the 
spectral condition:
\begin{equation}
|\omega| < \frac{1}{T}
.
\end{equation}
Otherwise the output waveform
will be distorted due to the
higher order terms in Eqs.~(\ref{eq:amp_imp}) and (\ref{eq:ph_imp}).
In electronic circuits, a rectangular pulse is most easily
generated.
But its spectrum has a long tail; $\sim 1/|\omega|$.
The tail must be suppressed with lowpass filters.
The cutoff frequency $1/T\sub{L}$ must be smaller than $1/T$.

\subsection{Lowpass filter for pulse preparation}

A simple method is to cascade suitable numbers of first-order lowpass
filters, whose transfer function is represented by Eq.~(\ref{eq:lowpass}),
as
\begin{equation}
 H\sub{L}^{(m)}(\omega)=\frac{1}{(1+\ii\omega \alpha_m T\sub{L})^m}
.
\end{equation}
where $\alpha_m=\sqrt{2^{1/m}-1}$ is a normalization 
parameter to keep the $3\,\U{dB}$
cutoff frequency constant.
It is reduced as $m$ is increased.
Otherwise, due to the decrease of bandwidth,
the pulse width is broadened and the pulse height is reduced.
For better low-pass characteristics, we can use
Bessel filters \cite{Tietze,Horowitz}, whose transfer function is given by
\begin{equation}
H\sub{B}^{(m)}(\omega)=\frac{1}{y_m(\ii\omega \alpha_m T\sub{L})}
\end{equation}
where $y_m(x)$ is the $m$-th Bessel polynomials
and the parameter $\alpha_m$ is determined so that
$H\sub{B}^{(m)}(T\sub{L}^{-1})=1/\sqrt{2}$.
\begin{figure}
\begin{center}
\includegraphics[scale=0.9]{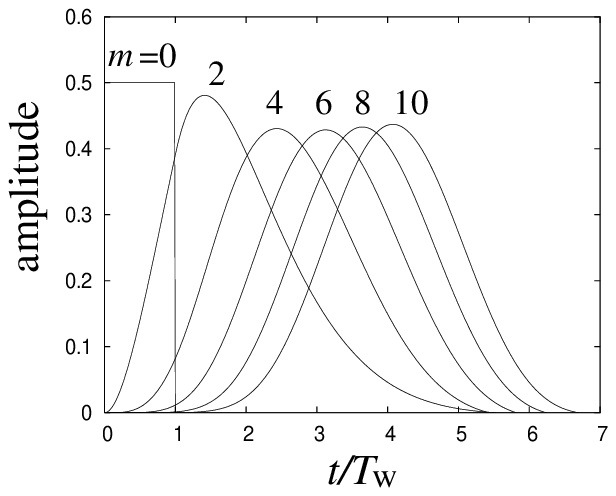}
\end{center}
\caption{
A rectangular pulse of width $T\sub{w} (=T\sub{L})$ is filtered
by $m$-th order Bessel filter with cutoff frequency
$1/T\sub{L}$.
The height of the original pulse ($m=0$) is unity
but it is halved in the graph.
}
\label{fig:filtering_time}
\end{figure}
\begin{figure}
\begin{center}
\includegraphics[scale=0.9]{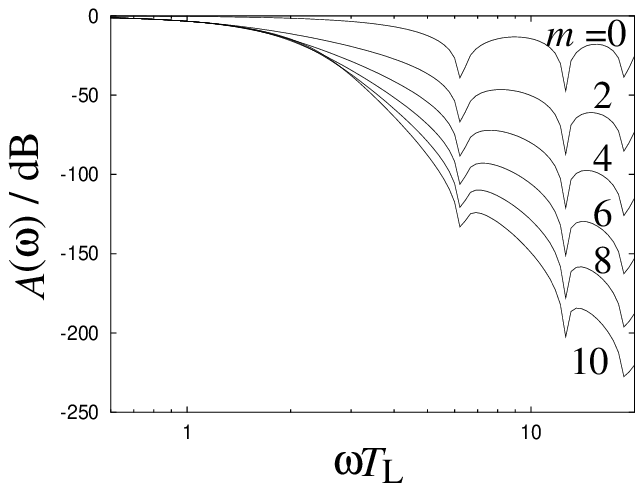}
\end{center}
\caption{Spectra of lowpass-filtered pulses (log-log plot).
The high frequency components strongly suppressed as
$m$ increases.}
\label{fig:filtering_freq}
\end{figure}
\begin{figure}
\begin{center}
\includegraphics[scale=0.9]{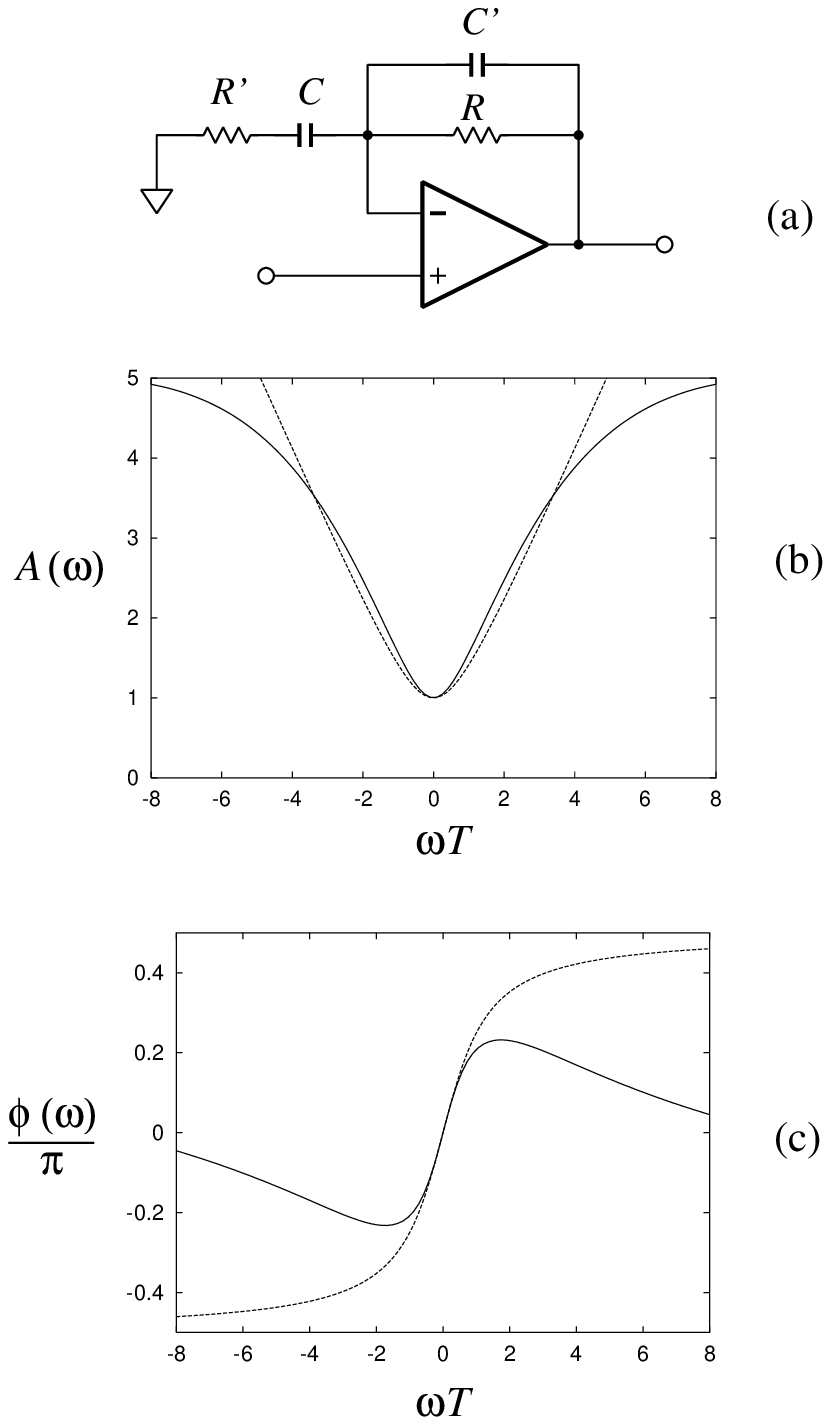}
\end{center}
\caption{
(a) Circuit diagram of a practical negative delay circuit.
Gain divergence for higher frequency  is suppressed by  $C'$ and $R'$ .
(b) Amplitude response function $A(\omega)$.
(c) Phase response function $\phi(\omega)$.
Parameters are $a=C'/C=0.2$, are $b=R'/R=0.05$.
The case of $a=b=0$ is also plotted.
Usable frequency region is limited within $|\omega|T < 1$.
}
\label{fig:practical_delay_cir}
\end{figure}

The effect of filtering is shown in 
Fig.~\ref{fig:filtering_time}
and Fig.~\ref{fig:filtering_freq}.
The initial, rectangular pulse 
$U(t)-U(t-T\sub{W})$ is 
send to a series of filters,
where $U(t)$ represent the unit step function.
As the order $m$ of the filter is increased, the high frequency tails are
more suppressed and
accordingly the waveform becomes smoother.
Exceeding a few stages, the waveforms look very similar
to each other.
But the leading edge scales $\sim t^{m}$ and
the peak position is delayed.
The delay, which is due to the phase transfer function of the
lowpass filters, is unavoidable.
We will see in Section VI that $m$ must be increased
in order to attain large advancement.
The pulse width approaches a value determined by
the cutoff frequency $T\sub{L}^{-1}$, if the initial pulse width 
is smaller than $T\sub{L}$.

From the way of preparation of input pulse with lowpass filters,
we can interpret that 
the leading edge is shaped so as to be more predictable.
The future can well be predicted, if enough
restrictions are imposed on the pulse.

\subsection{Gaussian pulse}
\label{sec:gauss}
The guassian pulse is widely used in negative delay or
superluminal propagation experiments.
It is because of the rapid tail-off of the spectrum and
its mathematical simplicity.
But there seems no natural implementation which generates a gaussian
pulse.
It is usually synthesized numerically and the calculated data is fed
to a digital-to-analog converter.
It should be noted that the ideal guassian pulse
has a infinitely long leading edge.
What we can generate practically is a truncated gaussian pulse:
\begin{equation}
v\sub{I}(t)=u(t)g(t-t_0)
,
\end{equation}
where $g(t)=\exp(-t^2/2T\sub{w}^2)$.
A causal function $u(t)$ is a unit step function or a smoothed
version of it.
This trancation necessarily introduces discontinuities
for $\dd^k v\sub{I}/\dd t^k$ $(k=0, 1, \ldots)$ at $t=0$ and
associates high frequency components.
However small they may look like,
eventually, they will be revealed 
by a negative delay circuit with advancement larger than $t_0$.
In order to reduce the effect of truncation, $t_0$ must be
increased, which corresponds to the increase of $m$ in the
pulse preparation with lowpass filters. 

\section{Experiment}
\subsection{Practical design}
Figure \ref{fig:practical_delay_cir}(a) shows 
a practical circuit for negative delays \cite{Nakanishi}.
The components $C'$ are $R'$ are added 
to the circuit of Fig.~\ref{fig:elementary_cir}(b) in order to 
suppress the gain for higher frequency.
Its transfer function can easily be derived.
First, we see that
\begin{equation}
 \tilde{V}_- =\frac{Z_C}{Z_C+Z_R}\tilde{V}\sub{out}
,
\end{equation}
where $Z_C=R'+1/\ii\omega C$,
$Z_R = (R^{-1}+\ii\omega C')^{-1}$.
If we assume a large gain of the operational amplifier,
the virtual short condition,
$\tilde{V}\sub{in}\sim \tilde{V}_-$, holds and we have
\begin{eqnarray}
H\sub{N}(\omega)&=& \tilde{V}\sub{out}/\tilde{V}\sub{in}
\nonumber\\
&=&(1+Z_R/Z_C)
\nonumber
\\
&=&
1+\frac{\ii\omega T}{(1+\ii\omega aT)(1+\ii\omega bT)}
,
\label{eq:real}
\end{eqnarray}
where $T=CR$.
If $a\equiv C'/C \ll 1$ and $b\equiv R'/R\ll 1$ are satisfied, then
the transfer function is approximated 
as $H\sub{N}(\omega)\sim 1+\ii\omega T$ near the origin ($|\omega| < 1/T$).
Thanks to $a$ or $b$,
the maximum gain is limited by $1/(a+b)$.
The response functions for $a=0.2$, $b=0.05$ are plotted 
in Figs.~\ref{fig:practical_delay_cir}(b) and (c).
The phase slope at the origin is almost conserved but
the usable bandwidth is reduced.

\begin{figure}
\begin{center}
\includegraphics[scale=0.9]{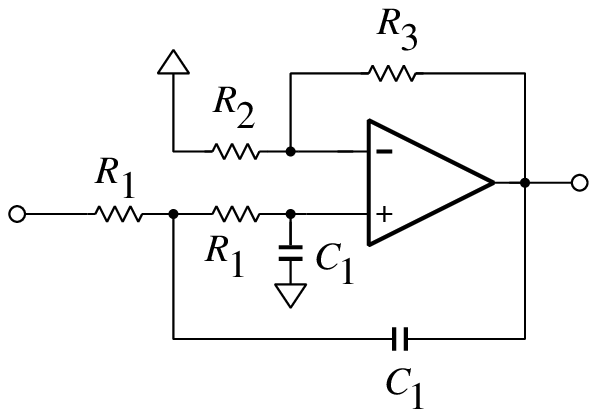}
\end{center}
\caption{2nd-order Bessel lowpass filter}
\label{fig:bessel}
\end{figure}
\begin{figure}
\begin{center}
\includegraphics[scale=0.9]{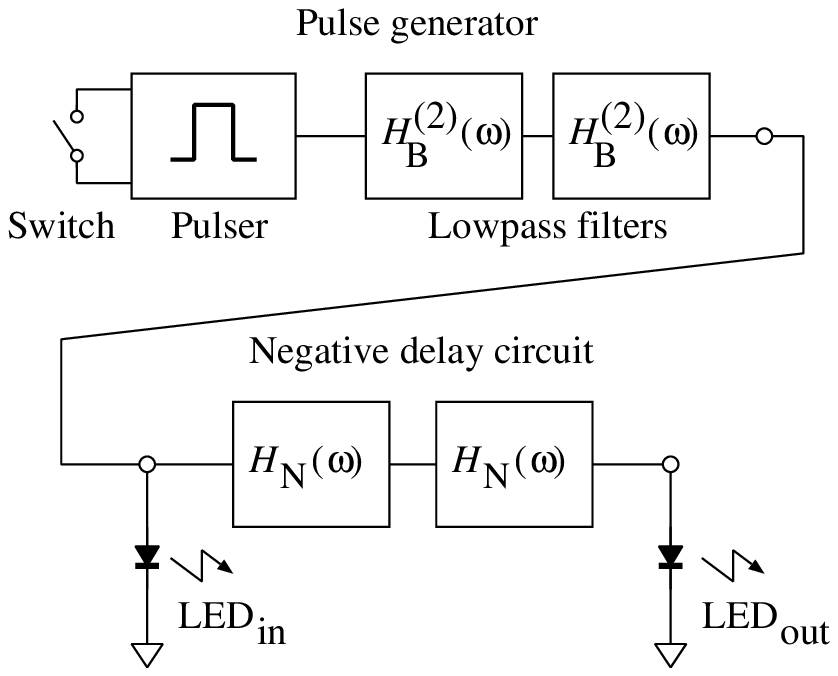}
\end{center}
\caption{Experimental setup}
\label{fig:block_diagram}
\end{figure}
Second-order Bessel filters are used in the experiment,
because the second-order filter can be realized 
with an operational amplifier \cite{Tietze,Horowitz}.
The circuit diagram is shown in Fig.~\ref{fig:bessel}.
The frequency response is
\begin{equation}
 H\sub{B}^{(2)} (\omega) = \frac{1}{1 + \ii\omega\alpha_2 T\sub{L}
 + (\ii \omega \alpha_2 T\sub{L})^2/3} 
,
\end{equation}
where 
$T\sub{L} = 1.272 R_1 C_1$ is the inverse of
cutoff frequency and $R_3/R_2 = 0.268$.

\subsection{Experimental result}

In Fig.~\ref{fig:block_diagram}, we show the overall block diagram for
the negative delay experiment.
The complete circuit diagram is presented in \cite{Nakanishi}.
The pulse generator on the top
is composed of a single-shot pulser
and two 2nd-order Bessel filters.
Triggered by the switch,
a timer IC (ICM7555) generates a rectangular pulse
with duration $1.5\,\U{s}$.
The pulse is shaped by the filters.
The cut-off frequency is chosen as
$\omega\sub{c}=1/T\sub{L}=0.35/T$, so that $A(\omega)$ and $\phi(\omega)$ can
be considered to be constant and linear, respectively.
$T=CR$ is the time constant of the negative delay circuits.

Two negative delay circuits are cascaded for larger advancement.
The circuit parameters are $R = 1\,\U{MO}$, $C = 0.22\,\U{uF}$,
$a=0.1$, $b=0.01$.

The input and output terminals are monitored by LEDs.
Their turn-on voltage is about $1.1\U{V}$.
\begin{figure}
\begin{center}
\includegraphics{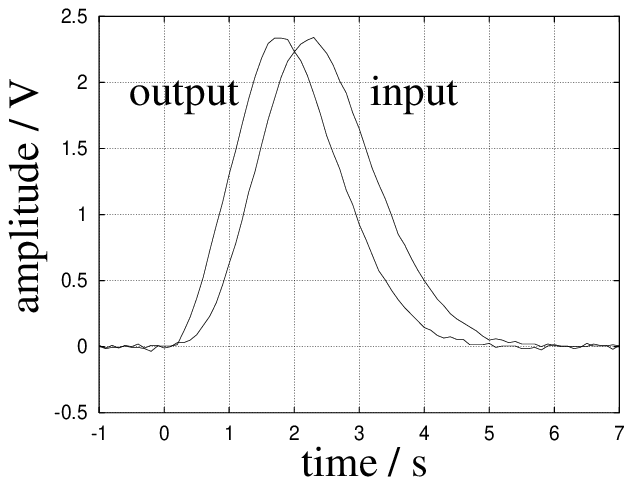}
\end{center}
\caption{Experimental result.
The traces of oscilloscope show the input and output
pulses.
Pulse is advanced about $0.5\,\U{s}$.
}
\label{fig:result}
\end{figure}

The experimental result is shown in Fig.~\ref{fig:result}.
The input and output waveforms are recorded with an oscilloscope.
The time origin ($t=0$) is the moment
when the switch is turned on or the rising edge of the
initial rectangular pulse.

We see that the output pulse precedes the input pulse considerably
(more than 20\% of the pulse width).
The slight distortion of the output waveform is caused by 
the non-ideal frequency dependence
of $A(\omega)$ and $\phi(\omega)$.

The observed advancement of $\sim 0.5\,\U{s}$
agrees well with the expected value $2T=2CR=0.44\,\U{s}$.

General purpose operational amplifiers (TL082) are used
for low-pass filters and negative delay circuits.
The time scale has been chosen so that we
can directly observe the negative delay with two LEDs connected at
the input and the output terminals. 
The whole experimental setup can be battery-operated and
self-contained.
If we want to observe the waveform, 
in stead of oscilloscopes, we can use two
analog voltmeters (or circuit testers) to monitor the waveforms.

\section{Cascading --- for larger advancement}

\subsection{Normalized advancement}
We have seen that in typical situations the advancement $(-t\sub{d})$
is fairly smaller than the pulse width $T\sub{w}$.
Typically the relative advancement $r\equiv |t\sub{d}|/T\sub{w}$ 
only reaches to a few percents.

To see the reason, we consider a gaussian pulse
$v(t)=V_0\exp(-t^2/2T\sub{w}^2)$ and its
Fourier transform $\tilde{V}(\omega)=\sqrt{2\pi}T\sub{w}
\exp(-\omega^2 T\sub{w}^2/2)$.
When it is passed through the negative delay circuit,
the power is amplified owing to 
the rising amplitude response (\ref{eq:hamp}).
We define the excess power gain $\eta$ as
\begin{equation}
1+\eta=\frac%
{\displaystyle\int_{-\infty}^{\infty} |H(\omega)\tilde{V}(\omega)|^2\dd\omega}%
{\displaystyle\int_{-\infty}^{\infty}|\tilde{V}(\omega)|^2\dd\omega}
,
\end{equation}
which can be used as a measure of distortion.
Now we have a relation between $\eta$ and the relative advancement $r$;
\begin{equation}
\eta=\frac%
{\displaystyle\int_{-\infty}^{\infty} 
  \omega^2 T^2 e^{-\omega^2 T\sub{w}^2/2} \dd\omega}%
{\displaystyle\int_{-\infty}^{\infty} e^{-\omega^2 T\sub{w}^2/2} \dd\omega}%
=2\left(\frac{T}{T\sub{w}}\right)^2=2r^2
.
\end{equation}
For example, if we allow $\eta=5\%$, then the 
relative advancement is $r=15\%$ at most.

If we want to increase the advancement $|t\sub{d}|$, 
for a given system, the bandwidth must be
reduced, which results in the increase of pulse width $T\sub{w}$
and $r$ does not increase.

\subsection{Degression of time constant}
We will try to increase the relative advancement 
by cascading the negative group delay circuits in series.
The transfer function for $n$ stages:
\begin{eqnarray}
H^{(n)}(\omega)&=&(1+\ii \omega T)^n 
,\\
A^{(n)}(\omega)&\sim&
1 + \frac{(\sqrt{n}\omega T)^2}{2},
\\
\phi^{(n)}(\omega)&\sim& n\omega T
.
\end{eqnarray}

At first, we may expect that 
the total advancement increases in proportional
to $n$ because the slope of $\phi^{(n)}(\omega)$ at the origin 
increases as $nT$.
But we should notice by looking at $A^{(n)}(\omega)$ that
the usable bandwidth is reduced as $n$ increased
for $T$ fixed.
In other words,
for a given input pulse width $T\sub{w}$,
$T$ must be reduced as $T\sub{w}/\sqrt{n}$.
Thus the total advancement scales as
\begin{equation}
(-t\sub{d}) \propto n\times(T\sub{w}/\sqrt{n})=\sqrt{n}T\sub{w}
.
\end{equation}

It should be noted that
the gain outside of the band is increases very rapidly.
Spectral tails of the input pulse must be suppressed enough.
Otherwise they could be amplified to distort the pulse shape.
From the asymptotic forms of $H^{(n)}(\omega)$ %
and $H^{(m)}\sub{L}(\omega)$,
we see that the condition $m>n$ must be satisfied.
Order $m$ of lowpass filters must be increased
cooperatively.
It is possible to increase the advancement as large as the pulse width
or more, but the advancement increases very slowly; $1/\sqrt{n}$.
\begin{figure}
\begin{center}
\includegraphics[scale=0.8]{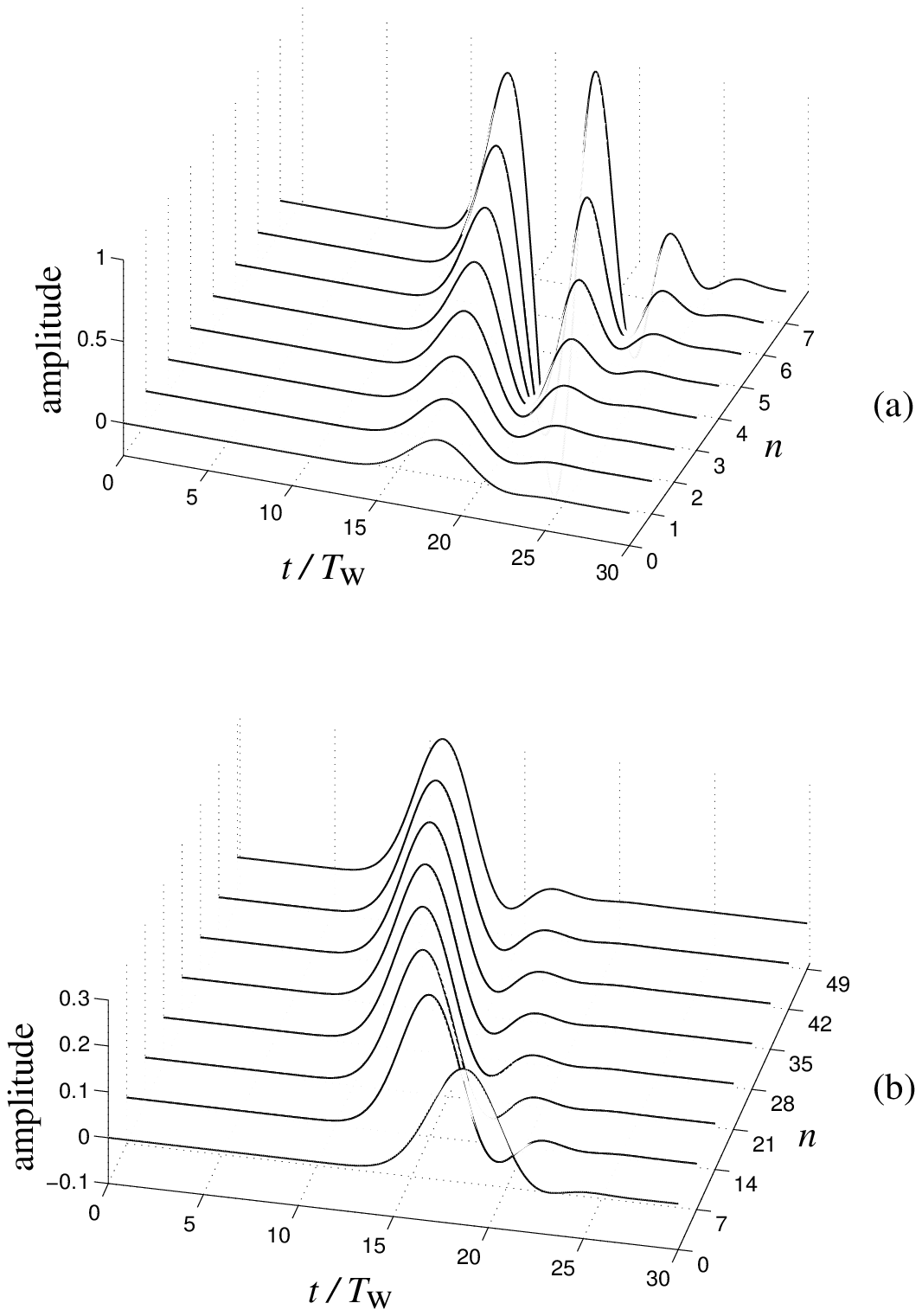}
\end{center}
\caption{
Cascading of negative delay circuits.
The input pulse is filtered by a series of
five 10th-order Bessel filters ($m=50$).
(a)
Simple cascading with fixed time constant.
The pulse is forwarded further as the number of circuits $n$ increases,
but the waveform is heavily distorted after a few stages.
(b)
Cascading with reduced time constant as $1/\sqrt{n}$.
Owing to small distortion we can increase $n$ even though
the advancement per stage is small.
With $n=49$ stages, the advancement clearly larger than the pulse width
is achieved.
}
\label{fig:cascading}
\end{figure}
Fig.~\ref{fig:cascading}(a) represents a example of
simple-minded cascading,
where $T$ is kept constant.
The waveforms are rapidly distorted as $n$ increased.

Fig.~\ref{fig:cascading}(b) shows the $1/\sqrt{n}$ cascading,
where $T$ is reduced as $T/\sqrt{n}$.
The pulse shape is preserved fairly well.
The input pulse is filtered with 
five 10-th order Bessel filters ($m=50$).
\begin{figure}
\begin{center}
\includegraphics[scale=0.75]{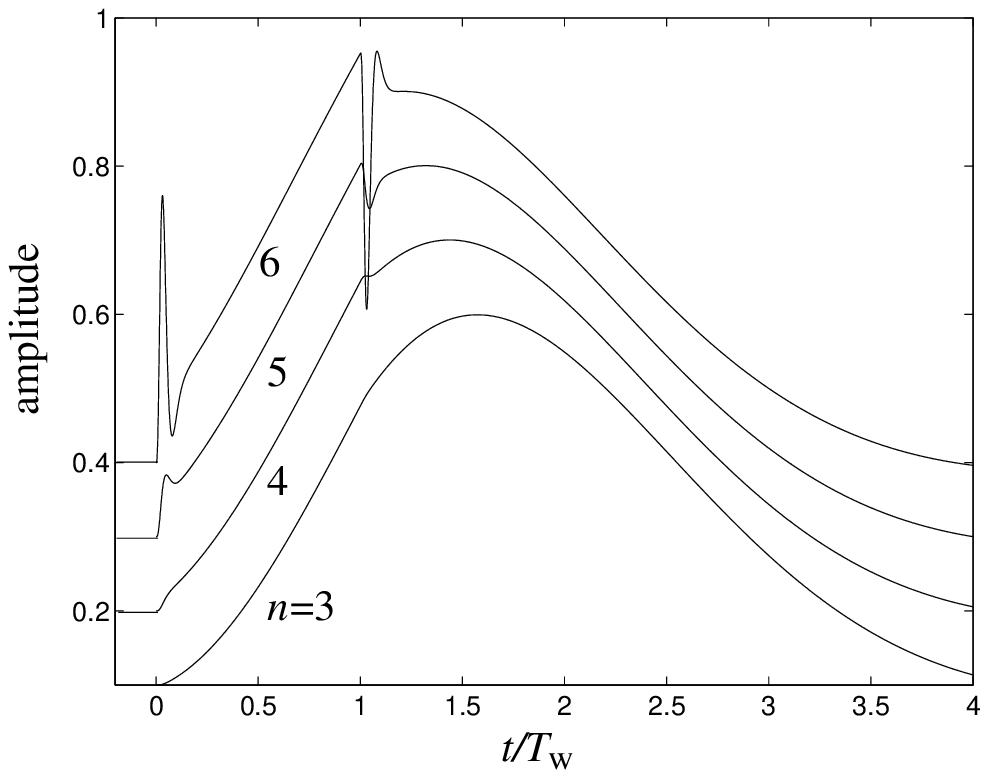}
\end{center}
\caption{Breakdown of cascading.
The input pulse is filtered by a 4-th order Bessel
filter ($m=4$).
We see that, for $n>m$,
the pulse advancement halts.
Especially at $t=0$ and $t=T\sub{w}$,
which correspond to the turn-on and the turn-off of
the original rectangular pulse, respectively,
strong deformation is observed.
Curves are offset for clarity.
}
\label{fig:excess_advancement}
\end{figure}
Figure \ref{fig:excess_advancement} shows the case of $n>m$.
The input rectangular pulse is filtered by a 4-th order Bessel
filter ($m=4$).
For $n=4$ cascading, scars of the initial pulse appear at
$t=0$ and $t=T\sub{w}$, and
they become more prominent as $n$ increases.

\subsection{Out-of-band gain}

As we have seen, the out-of-band gain 
is the primary obstacle toward large advancement.
In order to estimate how the gain increases we use
a realistic transfer function (\ref{eq:real}),
which has a finite maximum gain 
$A\sub{max}\sim 1+(a+b)^{-1}$.
For $a=0.2$, $b\ll a$, $n=50$, we have
\begin{equation}
 A\sub{max}^{n}=6^{50}\sim 10^{40}
.
\end{equation}
Such a huge gain will certainly induce instabilities.
The noise level must also be suppressed.
If we increase $a$ or $b$ to reduce $A\sub{max}$, then
the bandwidth is significantly reduced as seen in 
Fig.~\ref{fig:practical_delay_cir}.
For the reduced bandwidth, we have to increase the
pulse width $T\sub{w}$, which diminishes the relative advancement
$|t\sub{d}|/T\sub{w}$.

We see from this example, a large negative delay comparable
to the pulse width is very hard to achieve or almost prohibitive.
The allowable gain would be
limited by system-dependent factors such as 
a performance of active devices,
a threshold for instabilities, 
fluctuation due to quantum noise, and so on.

Again we notice the asymmetry between the negative and the
positive delays.
For positive delays, gain problem does not occur.
In fact, the cascading of lowpass filters yields a large amount of
delay without difficulty.

\section{Discussion}
\subsection{Interference in the time domain}

From the minimal transfer function, $H(\omega)=1+\ii\omega T$,
for the negative delay,
we see that in time domain the input-output relation
can be written as
\begin{equation}
 v\sub{out}(t)=\left(1+T\frac{\dd}{\dd t}\right)v\sub{in}(t)
= v\sub{in}(t)+T\frac{\dd v\sub{in}}{\dd t}(t)
.
\label{eq:derivative}
\end{equation}
Here the two terms interfere
constructively at the leading edge and
destructively at the trailing edge.
The addition of the time derivative to the original
pulse results in the pulse forwarding.

\begin{figure}
\begin{center}
\includegraphics[scale=0.8]{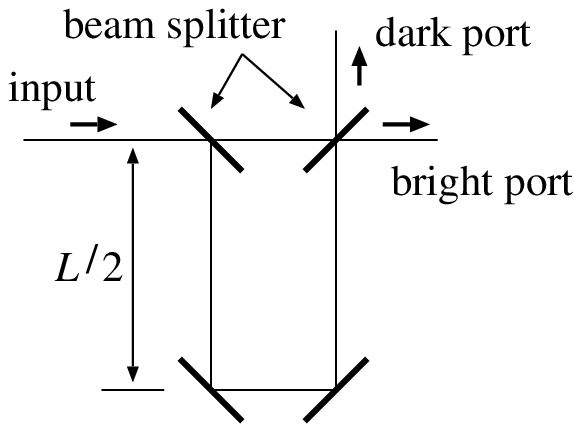}
\end{center}
\caption{
Mach-Zehnder interferometer as a negative group-delay device.
The reflectivity of two beam splitters is slightly smaller than
50\%.
The output from dark port is advanced.
}
\label{fig:mach_zehnder}
\end{figure}
\begin{figure}
\begin{center}
\includegraphics[scale=0.9]{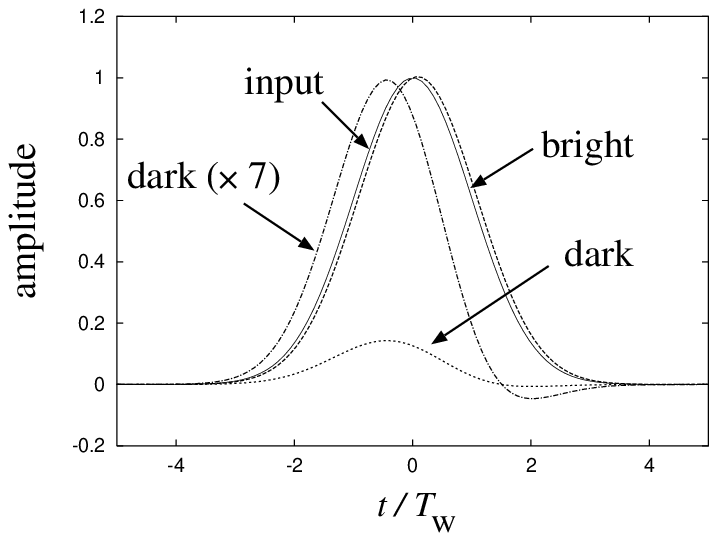}
\end{center}
\caption{
Pulse advancement by Mach-Zehnder interferometer.
A gaussian input pulse $g(t)=\exp(-t^2/2T\sub{w}^2)$ is
used.
The parameters are $\tau=0.17T\sub{w}$, and $\epsilon=0.06$.
}
\label{fig:interfero}
\end{figure}

This time domain picture is useful
to devise a new system which shows negative delays.
The Mach-Zehnder interferometer shown in 
Fig.~\ref{fig:mach_zehnder} is such an example.
First we assume $R=1/2$ for the reflectivity of the beam splitters.
The path difference $L$ is chosen so as to satisfy the
condition $\lambda\ll L = c\tau \ll cT\sub{w}$, 
where $\lambda$ is the wavelength,
$T\sub{w}$ is the pulse width,
and $\tau=L/c$ is the delay time due to the path difference.
We can tune the path length so that the transmission for
one port is unity.
Then for the steady state, there appears no output 
at the other port (dark port) owing to the destructive interference.

For time dependent inputs, however, the cancellation is
incomplete and the output corresponding to the time derivative
of the amplitude appears at the dark port (see Fig.~\ref{fig:interfero}).
If we superpose this output with the original waveform,
we will have the advancement as shown in Eq.~(\ref{eq:derivative}).
The superposition can easily be provided by
unbalancing the amplitude of each path.
We set the reflectivity of the two beam splitters is slightly smaller
than 50\% : $R=1/2-\epsilon$.
Then the output of the dark port becomes
\begin{eqnarray}
\tilde{E}\sub{dark}(t)
&=&(1-R) \tilde{E}\sub{in}(t)
-R\tilde{E}\sub{in}(t-\tau)
\nonumber
\\
&\sim&
2\epsilon\left(1+\frac{\tau}{4\epsilon}\frac{\dd}{\dd t}\right)
\tilde{E}\sub{in}(t)
,
\end{eqnarray}
and the advancement of $\tau/4\epsilon$ is achieved.
$\tilde{E}\sub{in}(t)$ and $\tilde{E}\sub{dark}(t)$ are
the envelope of the input field and the dark port field,
respectively.
The usable bandwidth of the system is $\Delta\omega \sim 1/\tau$,
for which the darkness of the dark port is ensured.

This is an example of all-passive systems with negative delay.
It should be noted that when we increase 
the advancement by decreasing $\epsilon$, the transmission
is decreased accordingly.
It is also true for the superluminal propagation
of evanescent waves and tunneling waves \cite{Enders,Steinberg}.

This model convinces us that the negative group delay and the
superluminal group velocity are the simple consequence of 
wave interference.

\subsection{Causality and negative delay}
 \begin{figure}
\begin{center}
\includegraphics[scale=0.8]{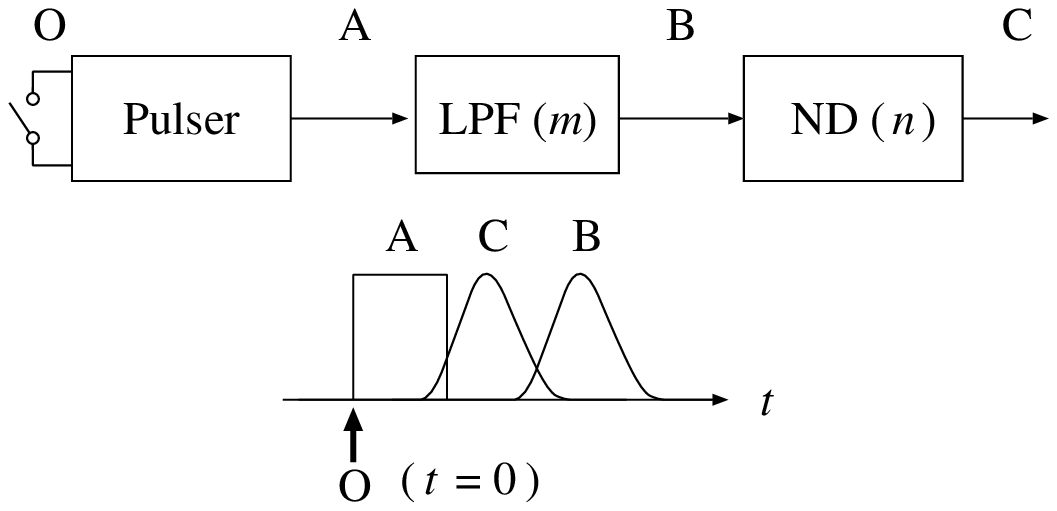}
\end{center}
\caption{
Communication channel with negative group delay.
Inset shows timing of pulses at each site.
}
\label{fig:causality}
 \end{figure}

It has been well recognized and confirmed in many ways that
the front velocity is
connected with the causality,
and the causality has no
direct connection with the group velocity.
But one is still apt to connect the group velocity with
the causality because 
many practical communication systems utilize pulse 
modulation to send information.

Let us examine our system shown in 
Fig.~\ref{fig:causality} in terms of causality.
By pushing the switch (O), a rectangular pulse (A) is generated.
Feeding it into the lowpass filter,
a band-limited pulse (B) is prepared.
Finally sending it through the negative delay circuit,
an advanced pulse (C) is created.
This would be a cause-effect chain in a casual sense.
However the reversal of the chronological order 
between (B) and (C) causes the trouble
for the naive picture.
One may think that making use of this twist it is possible to
send information to the past despite of the causality.
Of course this is wrong.

First we should realize that in a strict sense
all the pulses (A), (B), and (C) are
causal to each other because they are the signals in a single lumped
system.
When an impulse is applied to a quiescent lumped system,
then all parts respond instantaneously.
Some of them look delayed, but they just started smoothly
as $t^k$ ($k>1$).
All the pulse fronts share the time $t=0$, 
when the switch is turned on (O).
Therefore the above discussion on the order is totally pointless.

But one may still shelve the theory, which deals with
the almost unseen signals just after $t=0$, considering practical
situation where the information is related to the peak position
or the rising edge where a half of the pulse peak is reached.
However, in order to generate a smooth pulse (B) which is acceptable
to the negative delay circuit, we have to make a decision
well in advance (before $t=0$ in this case) because of the delay
caused by the lowpass filter.
Once we miss the timing, the number $m$ of the lowpass filters must
be reduced in order to catch up.
But breaking the condition $m>n$, the pulse cannot be forwarded
any more and is distorted badly.

Let us regard the negative delay circuit of 
Fig.~\ref{fig:causality} 
as a communication channel.
We assign three people, Alice, Bob, and Clare, on 
the sites (A), (B), and (C), respectively.
Clare always finds a pulse before Bob does, i.e.,
she can always predict Bob's pulse.
But Bob has no control over his pulses;
he cannot cancel a pulse initiated by Alice.
The real sender of the pulse is not Bob, but Alice. 
Bob is just an observer standing at the sending site.
This scenario tells us that comparing the input and the output
pulses of superluminal channel is somewhat nonsense and
that the real start point of the input pulse should be
considered (See Sec. \ref{sec:gauss}).

\section{Concluding remarks}

The negative group delay is already utilized
in many practical applications implicitly.
Signals from slow sensors, such as a hotwire 
anemometer, are compensated by a differentiator
with a transfer function $1+\ii\omega T$.
In PID (Proportional, Integral, and Derivative) controllers
the derivative element (D) is used to predict the behavior of
the system and to improve the dynamical response.
When a capacitive load is connected to the output
of an operational amplifier, an additional feedback loop
with derivative element is used, which is called 
lead compensations.
All these efforts are to compensate delays in a system
as far as possible but 
the excessive use will result in instabilities or noise problems.

We have explored many aspects of negative delays and superluminality
utilizing circuit models.
The use of circuit model is very helpful because 
the choice of parameters are very flexible and
many handy circuit-simulation softwares are available.
Extension to nonlinear cases and to distributed
systems \cite{Ziolkowski} will be very interesting.

\section*{acknowledgments}
One of the authors (MK) greatly appreciates
stimulating and inspiring discussions with R. Chiao and 
all the participants of the mini program
on Quantum Optics at the Institute of Theoretical
Physics, University of California Santa Barbara.
He also thanks K. Shimoda for informing about practical use of
negative delay circuits.

\end{document}